\documentclass{moriond}

\def\Journal#1#2#3#4{{#1} {\bf #2}, #3 (#4)}

\def\NPB{{\em Nucl. Phys.} B}
\def\PLB{{\em Phys. Lett.}  B}
\def\PRL{\em Phys. Rev. Lett.}
\def\PRD{{\em Phys. Rev.} D}

\def\be{\begin{equation}}
\def\ee{\end{equation}}
\def\bea{\begin{eqnarray}}
\def\eea{\end{eqnarray}}

\newcommand{\Photo}{}
\usepackage{ifthen} 
\newboolean{uprightparticles}
\setboolean{uprightparticles}{false} 
\usepackage{xspace} 
\usepackage{upgreek}

\def\lhcb   {\mbox{LHCb}\xspace}

\def\babar  {\mbox{BaBar}\xspace}
\def\belle  {\mbox{Belle}\xspace}

\def\besiii {\mbox{BESIII}\xspace}
\def\cleo   {\mbox{CLEO}\xspace}

\def\MagUp {\mbox{\em Mag\kern -0.05em Up}\xspace}

\ifthenelse{\boolean{uprightparticles}}%
{

 \def\Ppi         {\ensuremath{\uppi}\xspace}

 \def\PDelta      {\ensuremath{\Delta}\xspace}                 
 \def\PXi         {\ensuremath{\Xi}\xspace}                 
 \def\PLambda     {\ensuremath{\Lambda}\xspace}                 
 \def\PSigma      {\ensuremath{\Sigma}\xspace}                 
 \def\POmega      {\ensuremath{\Omega}\xspace}                 
 \def\PUpsilon    {\ensuremath{\Upsilon}\xspace}

 \def\PB      {\ensuremath{\mathrm{B}}\xspace}                 
                  
 \def\PD      {\ensuremath{\mathrm{D}}\xspace}

 \def\PK      {\ensuremath{\mathrm{K}}\xspace}

 \def\Pb      {\ensuremath{\mathrm{b}}\xspace}                 
 \def\Pc      {\ensuremath{\mathrm{c}}\xspace}                 
 \def\Pd      {\ensuremath{\mathrm{d}}\xspace}

 \def\Pi      {\ensuremath{\mathrm{i}}\xspace}

 \def\Pp      {\ensuremath{\mathrm{p}}\xspace}

 \def\Ps      {\ensuremath{\mathrm{s}}\xspace}                 
 \def\Pt      {\ensuremath{\mathrm{t}}\xspace}                 
 \def\Pu      {\ensuremath{\mathrm{u}}\xspace}

 \def\thebaroffset{0.0em}
}
{

 \def\Ppi         {\ensuremath{\pi}\xspace}

 \mathchardef\PDelta="7101
 \mathchardef\PXi="7104
 \mathchardef\PLambda="7103
 \mathchardef\PSigma="7106
 \mathchardef\POmega="710A
 \mathchardef\PUpsilon="7107
                  
 \def\PB      {\ensuremath{B}\xspace}                 
                  
 \def\PD      {\ensuremath{D}\xspace}

 \def\PK      {\ensuremath{K}\xspace}

 \def\Pb      {\ensuremath{b}\xspace}                 
 \def\Pc      {\ensuremath{c}\xspace}                 
 \def\Pd      {\ensuremath{d}\xspace}

 \def\Pi      {\ensuremath{i}\xspace}

 \def\Pp      {\ensuremath{p}\xspace}

 \def\Ps      {\ensuremath{s}\xspace}                 
 \def\Pt      {\ensuremath{t}\xspace}                 
 \def\Pu      {\ensuremath{u}\xspace}

 \def\thebaroffset{0.18em}
}
\newcommand{\offsetoverline}[2][\thebaroffset]{\kern #1\overline{\kern -#1 #2}}%

\makeatletter
\ifcase \@ptsize \relax
  \newcommand{\miniscule}{\@setfontsize\miniscule{4}{5}}
\or
  \newcommand{\miniscule}{\@setfontsize\miniscule{5}{6}}
\or
  \newcommand{\miniscule}{\@setfontsize\miniscule{5}{6}}
\fi
\makeatother

\DeclareRobustCommand{\optbar}[1]{\shortstack{{\miniscule (\rule[.5ex]{1.25em}{.18mm})}
  \\ [-.7ex] $#1$}}

\def\uquark    {{\ensuremath{\Pu}}\xspace}

\def\dquark    {{\ensuremath{\Pd}}\xspace}

\def\squark    {{\ensuremath{\Ps}}\xspace}
\def\squarkbar {{\ensuremath{\overline \squark}}\xspace}

\def\cquark    {{\ensuremath{\Pc}}\xspace}
\def\cquarkbar {{\ensuremath{\overline \cquark}}\xspace}

\def\bquark    {{\ensuremath{\Pb}}\xspace}
\def\bquarkbar {{\ensuremath{\overline \bquark}}\xspace}

\def\tquark    {{\ensuremath{\Pt}}\xspace}

\def\pion   {{\ensuremath{\Ppi}}\xspace}

\def\pip    {{\ensuremath{\pion^+}}\xspace}
\def\pim    {{\ensuremath{\pion^-}}\xspace}
\def\pipm   {{\ensuremath{\pion^\pm}}\xspace}
\def\pimp   {{\ensuremath{\pion^\mp}}\xspace}

\def\kaon    {{\ensuremath{\PK}}\xspace}

\def\KorKbar {\kern \thebaroffset\optbar{\kern -\thebaroffset \PK}{}\xspace}

\def\Kp      {{\ensuremath{\kaon^+}}\xspace}
\def\Km      {{\ensuremath{\kaon^-}}\xspace}

\def\Kmp     {{\ensuremath{\kaon^\mp}}\xspace}

\def\Dbar    {{\ensuremath{\offsetoverline{\PD}}}\xspace}
\def\D       {{\ensuremath{\PD}}\xspace}

\def\DorDbar {\kern \thebaroffset\optbar{\kern -\thebaroffset \PD}\xspace}
\def\Dz      {{\ensuremath{\D^0}}\xspace}
\def\Dzb     {{\ensuremath{\Dbar{}^0}}\xspace}
\def\Dp      {{\ensuremath{\D^+}}\xspace}
\def\Dm      {{\ensuremath{\D^-}}\xspace}

\def\DpDm    {\ensuremath{\Dp {\kern -0.16em \Dm}}\xspace}

\def\Dstarp  {{\ensuremath{\D^{*+}}}\xspace}
\def\Dstarm  {{\ensuremath{\D^{*-}}}\xspace}

\def\Ds      {{\ensuremath{\D^+_\squark}}\xspace}
\def\Dsp     {{\ensuremath{\D^+_\squark}}\xspace}

\def\DporDsp {{\ensuremath{\D_{(\squark)}^+}}\xspace}
\def\DmorDsm {{\ensuremath{\D{}_{(\squark)}^-}}\xspace}

\def\B       {{\ensuremath{\PB}}\xspace}

\def\BorBbar {\kern \thebaroffset\optbar{\kern -\thebaroffset \PB}\xspace}

\def\Bd      {{\ensuremath{\B^0}}\xspace}

\def\BdorBdbar {\kern \thebaroffset\optbar{\kern -\thebaroffset \Bd}\xspace}

\def\Bs      {{\ensuremath{\B^0_\squark}}\xspace}

\def\BsorBsbar {\kern \thebaroffset\optbar{\kern -\thebaroffset \Bs}\xspace}

\def\Y#1S{\ensuremath{\PUpsilon{(#1S)}}\xspace}

\def\proton      {{\ensuremath{\Pp}}\xspace}

\def\LorLbar     {\kern \thebaroffset\optbar{\kern -\thebaroffset \PLambda}\xspace}

\def\to                 {\ensuremath{\rightarrow}\xspace}

\def\CP                {{\ensuremath{C\!P}}\xspace}

\def\Vud  {{\ensuremath{V_{\uquark\dquark}^{\phantom{\ast}}}}\xspace}
\def\Vcd  {{\ensuremath{V_{\cquark\dquark}^{\phantom{\ast}}}}\xspace}
\def\Vtd  {{\ensuremath{V_{\tquark\dquark}^{\phantom{\ast}}}}\xspace}

\def\Vubs  {{\ensuremath{V_{\uquark\bquark}^\ast}}\xspace}
\def\Vcbs  {{\ensuremath{V_{\cquark\bquark}^\ast}}\xspace}
\def\Vtbs  {{\ensuremath{V_{\tquark\bquark}^\ast}}\xspace}

\def\AT#1     {\ensuremath{A_{\mathrm{T}}^{#1}}\xspace}           

\def\C#1      {\ensuremath{\mathcal{C}_{#1}}\xspace}                       
\def\Cp#1     {\ensuremath{\mathcal{C}_{#1}^{'}}\xspace}                    
\def\Ceff#1   {\ensuremath{\mathcal{C}_{#1}^{\mathrm{(eff)}}}\xspace}        
\def\Cpeff#1  {\ensuremath{\mathcal{C}_{#1}^{'\mathrm{(eff)}}}\xspace}       
\def\Ope#1    {\ensuremath{\mathcal{O}_{#1}}\xspace}                       
\def\Opep#1   {\ensuremath{\mathcal{O}_{#1}^{'}}\xspace}                    

\newcommand{\aunit}[1]{\ensuremath{\text{\,#1}}}       

\newcommand{\tev}{\aunit{Te\kern -0.1em V}\xspace}
\newcommand{\gev}{\aunit{Ge\kern -0.1em V}\xspace}
\newcommand{\mev}{\aunit{Me\kern -0.1em V}\xspace}
\newcommand{\kev}{\aunit{ke\kern -0.1em V}\xspace}
\newcommand{\ev}{\aunit{e\kern -0.1em V}\xspace}
 
\newcommand{\mevc}{\ensuremath{\aunit{Me\kern -0.1em V\!/}c}\xspace}
\newcommand{\gevc}{\ensuremath{\aunit{Ge\kern -0.1em V\!/}c}\xspace}
\newcommand{\mevcc}{\ensuremath{\aunit{Me\kern -0.1em V\!/}c^2}\xspace}
\newcommand{\gevcc}{\ensuremath{\aunit{Ge\kern -0.1em V\!/}c^2}\xspace}

\def\gsim{{~\raise.15em\hbox{$>$}\kern-.85em
          \lower.35em\hbox{$\sim$}~}\xspace}
\def\lsim{{~\raise.15em\hbox{$<$}\kern-.85em
          \lower.35em\hbox{$\sim$}~}\xspace}

\def\tell1  {TELL1\xspace}
\def\ukl1   {UKL1\xspace}

\usepackage{lineno}

\begin{document}
\vspace*{4cm}
\title{\CP violation in beauty and charm quarks at \lhcb}

\author{ D. Manuzzi, on behalf of the \lhcb collaboration }

\address{Departement of Physics, University of Bologna, \\via Irnerio 46, Bologna, Italy}

\maketitle

\abstracts{
The \lhcb experiment has been reporting remarkable \CP-violation (CPV) measurements concerning the sectors of \bquark and \cquark quarks.
Recently, the new measurements of time-integrated CPV with \mbox{$\Dz\to\Kp\Km$} decays led to the first evidence ($3.8\sigma$) of CPV in a single charmed decay.
The first search for CPV in the \mbox{$\DporDsp\to \Kp\Kp\Km$} decays was executed.
The knowledge of the CKM parameter $\gamma$ was improved by new results; 
the current combination of the \lhcb measurements is: $\left(63.8^{+3.5}_{-3.7}\right)^\circ$.
The study of \mbox{$\Bs\to\phi\phi$} decays resulted in the most precise measurement of time-dependent CPV in any penguin-dominated \B meson decay. 
All the results are consistent with the Standard-Model predictions.
}
\section{Introduction}
The \lhcb experiment is a single-arm forward spectrometer operating at the Large Hadron Collider (LHC) at CERN\cite{LHCbLHC}.
The \lhcb design is particularly effective for the indirect search for new physics through the study of \CP violation (CPV) in heavy mesons. 
Thanks to its peculiar pseudorapidity acceptance ($\eta\in [1.6,49]$), \lhcb benefits of a relatively high production cross-sections for $\bquark\bquarkbar$ and $\cquark\cquarkbar$ pairs\cite{LHCbXsections} (see Table~\ref{tab:crossSection_Lumi}). 
Other relevant features are the excellent decay-time resolution for \B mesons ($\approx 40~{\rm fs}$), the momentum resolution (\mbox{$\delta p/p \sim 0.5\% - 0.8\%$}), and the particle identification performance\cite{LHCbPerformance}.
This proceeding documents the results that \lhcb has obtained in the CPV sector of \bquark and \cquark quarks over the last six moths.

\begin{table}[h!]
\caption[]{General characteristics of \lhcb data-taking periods. From left to right, the columns of the table report: the number of the run, the first an the last data-taking year of the period, the center-of-mass energy of the \proton\proton collision provided by the LHC, the production cross-section for \bquark\bquarkbar and \cquark\cquarkbar pairs~\cite{LHCbXsections}, and the integrated luminosity.}
\label{tab:crossSection_Lumi}
\vspace{0.4cm}
\centering
\begin{tabular}{|c|c|c|c|c|c|}
\hline
Run & Years  & $\sqrt{s}~[{\rm TeV}]$ & $\sigma(pp\to\bquark\bquarkbar X)~[\mu{\rm b}]$ & $\sigma(pp\to\cquark\cquarkbar X)~[\mu{\rm b}]$ & $\mathcal{L}_{\rm int}~[{\rm fb}^{-1}]$ \\ 
\hline
1  & 2011-2012 & $7$ & $72.0\pm 6.8$ & $1419\pm 134$ & $3$ \\
2  & 2015-2018 & $13$ & $144\pm 21$ & $2369\pm 192$ & $9$ \\
\hline
\end{tabular}
\end{table}


\section{\CP violation in the charm sector}
Up to now the unique observation of CPV in the charm sector is\cite{DeltaACP}:
{$\Delta A_\CP \equiv A_\CP(\Dz\to \Kp\Km)-A_\CP(\Dz\to \pip\pim) = (-1.54 \pm  0.29)\times 10^{-3}$},
where {$A_\CP(\Dz \to f) = [\Gamma(\Dz\to f)- \Gamma(\Dzb\to f)]/[\Gamma(\Dz\to f)+ \Gamma(\Dzb\to f)]$}
is the time-integrated \CP asymmetry for the mode involving the final state $f$.
The measured value generated a debate about its interpretation\cite{DetlaACP_debate}, because it lays close to the edge of the range predicted\cite{DetlaACP_debate} by the Standard Model (SM). 
To solve the puzzle, further measurements are required.
The $\Delta A_\CP$ observable relies on the difference between the \CP asymmetries of two channels to cancel out experimental biases (nuisance asymmetries). 
More recently, \lhcb published the direct measurement of  \mbox{$A_\CP(\Dz\to \Kp\Km)$} with Run2 data.
The analysis\cite{ACPKK} exploits the \mbox{$\Dstarp \to (\Dz\to \Kp\Km) \pip$} decay chain, where the charge of the final state pion tags the flavour of the neutral \D meson at production. 
An invariant-mass fit determines the raw asymmetry between the yields of the candidates reconstructed as \Dz or \Dzb: \mbox{$A(\Dz) = [N(\Dz)-N(\Dzb)]/ [N(\Dz)-N(\Dzb)]$}.
This quantity is the sum of multiple contributions: \mbox{$A(\Dz) = A_{\CP}(\Dz\to \Kp\Km) + A_{\rm prod}(\Dz) + A_{\rm det}(\pip)$},
namely the physical \CP asymmetry, the production asymmetry due to the difference between the production cross-sections of \Dstarp and \Dstarm mesons, and the detection asymmetry due to the different reconstruction efficiency of final states particles. 
To extract the $A_{\CP}$, raw asymmetries of various Cabibbo-favoured modes are measured and combined\footnote{ 
A first method uses: \mbox{$D^{*+}\to (D^0\to K^-\pi^+) \pi^+$}, \mbox{$D^+\to K^-\pi^+\pi^+$}, and \mbox{$D^+\to \overline K^0\pi^+$} decays; a second method exploits: \mbox{$D^{*+}\to (D^0\to K^-\pi^+) \pi^+$}, \mbox{$D^+_s\to \phi\pi^+$}, and \mbox{$D^+_s\to \overline K^0K^+$} decays.}.
The procedure also requires accurate kinematic reweighting of all the channels.
The final result is\cite{ACPKK}
\mbox{$A_{\CP}(D^0\to K^+K^-)  = (6.8 \pm 5.4 \pm 1.6)\times 10^{-4}$},
where the uncertainties are statistical and systematic, respectively. The total uncertainty is about a half the uncertainty of the previous world average\cite{HFLAV}. 
Up to first order in the \Dz mixing parameters, the following relation holds\cite{ACPcharm}: \mbox{$A_{\CP}(\D\to f) \simeq a_f^d + \Delta Y_f \cdot (\langle t \rangle_f / \tau_D)$}, where $a^d_f$ quantifies the CPV in the decay, $\Delta Y_f$ is related to the presence of mixing-induced CPV.
$\tau_D$ is the lifetime of the \Dz meson, and $\langle t \rangle_f$ is the mean decay time of the \Dz candidates in the data sample. 
Combining \lhcb results and world averages\cite{DeltaACP,ACPKK,AKKextInputs,PDG2022}, one obtains: $a_{\Kp\Km}^d = (7.7 \pm 5.7)\times 10^{-4}$ and $a_{\pip\pim}^d = (23.2 \pm 6.1)\times 10^{-4}$.
In particular, $a_{\pip\pim}^d$ is different from zero at $3.8\sigma$ level. 
This is the first evidence of CPV in a single charmed decay mode. 

\begin{figure}[t]
\begin{minipage}{0.49\linewidth}
\centerline{\includegraphics[width=0.60\linewidth]{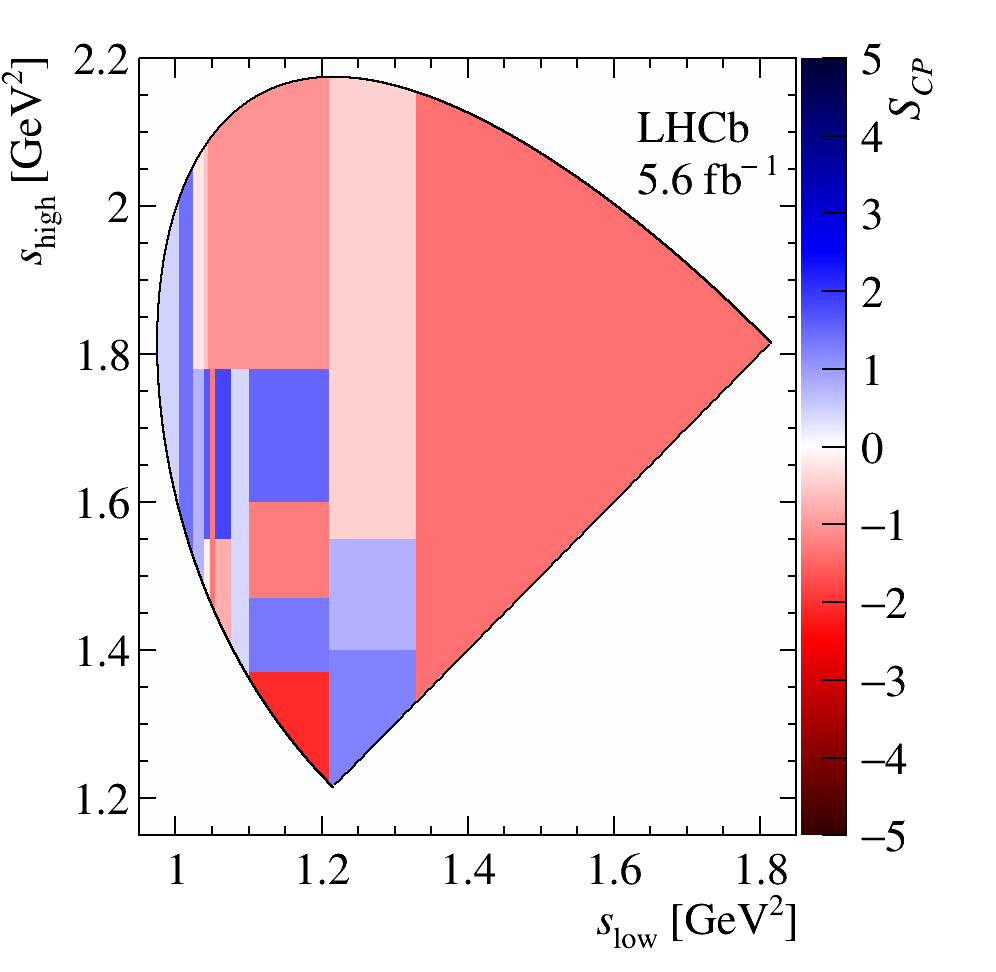}}
\end{minipage}
\begin{minipage}{0.49\linewidth}
\centerline{\includegraphics[width=0.60\linewidth]{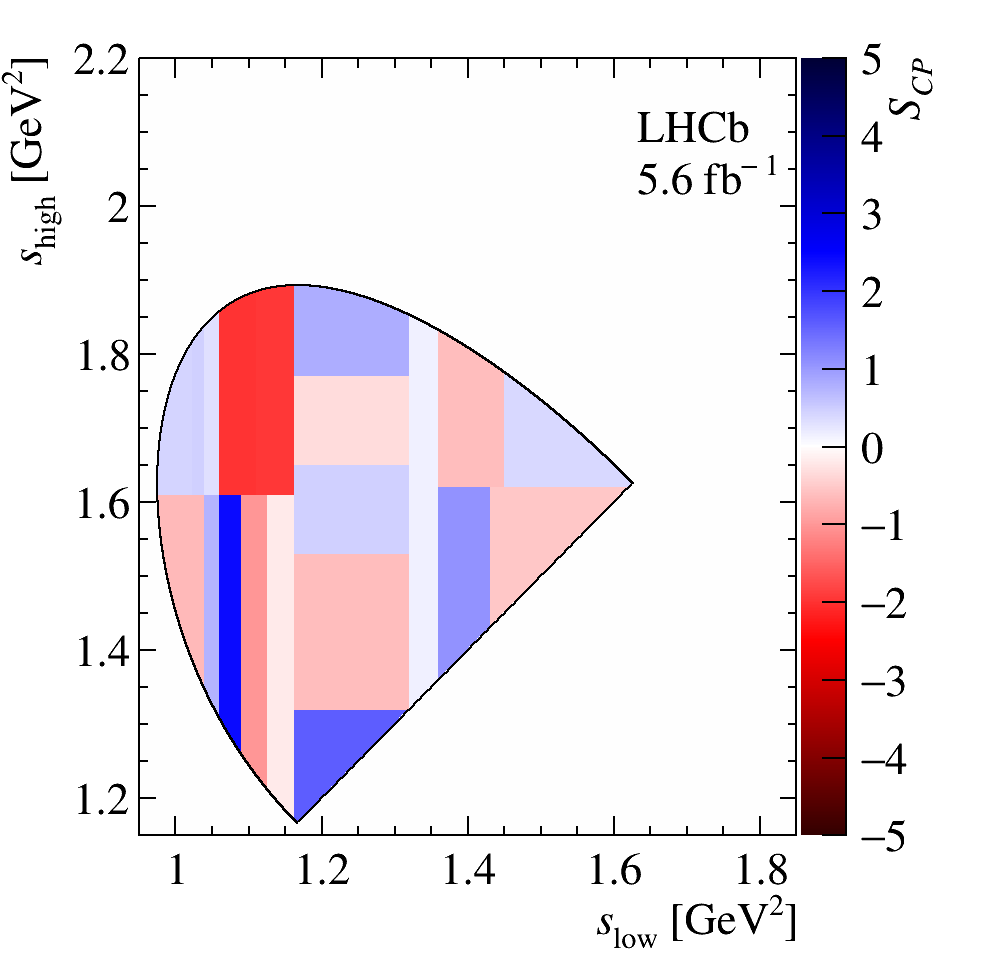}}
\end{minipage}
\caption[]{$\mathcal{S}_\CP$ values across the Dalitz plot for (left) $\Ds\to\Km\Kp\Kp$ and (right) $\Dp \to \Km\Kp\Kp$ candidates\cite{CPV_D2KKK}.}
\label{fig:aCPV_D2KKK}
\end{figure}
In multibody decays the CPV depends on the phase space. Hence, local \CP asymmetries may be larger than the integrated ones.
A good candidate for this search is the Cabibbo-suppressed \mbox{$\Dsp \to \Kp\Kp\Km$} decay. 
The companion doubly-Cabibbo-suppressed \mbox{$\Dp \to \Kp\Kp\Km$} decay is relevant, as well. 
Since the SM suppresses all CPV effects in this mode, an eventual observation would point towards new physics.
\lhcb has studied these cases with Run2 data. 
The analysis\cite{CPV_D2KKK} splits the Dalitz plots of the two channels in 21 bins.
Their definition reproduce the pattern of the main resonances to enhance the sensitivity.
An invariant-mass fit is performed in each bin to get the yields, $N^i$, of \DporDsp and \DmorDsm candidates.
The following \CP-related observable is calculated:
$\mathcal{S}_\CP^i = [N^i(\DporDsp) - \alpha N^i(\DmorDsm)] /
[\alpha ( \delta^2_{N^i(\DporDsp)} + \delta^2_{N^i(\DmorDsm)})]^{\frac{1}{2}}$
 with  $\alpha = \sum_i N^i(\DporDsp)/\sum_i N^i(\DmorDsm)$,
where $\delta_{N^i(\D)}$ is the statistical uncertainty of the yield at subscript.
Global nuisance asymmetries do not affect $\mathcal{S}_\CP^i$~\cite{Miranda}.
Local changes in the nuisance asymmetries are verified to be negligible using simulation and control samples of Cabibbo-Favoured decays\footnote{In particular \mbox{$\Dp \to \Km\pip\pip$} and \mbox{$\Dp \to \Km\Kp\pip$} decays.}.
Figure~\ref{fig:aCPV_D2KKK} illustrates the measured $\mathcal{S}^i_\CP$ values.
As a global test against the hypothesis of null CPV, the \mbox{$\chi^2 = \sum_i \left(\mathcal{S}_\CP^i\right)^2$} observable is computed.
The corresponding $p$-values are $31.6\%$ in the \Dp case and $13.3\%$ in the  \Dsp case. 
Hence, no CPV evidence is found. 
This is the first CPV search in \mbox{$\DporDsp \to \Kp\Kp\Km$} decays.
\section{Direct measurements of the CKM parameter $\gamma$}
The SM assumes the unitarity of the CKM matrix\cite{CKM}. 
This condition implies various relations among its elements. 
A relevant one is \mbox{$\Vud\Vubs + \Vcd\Vcbs + \Vtd\Vtbs = 0$}.
In the complex plane, it defines the notorious Unitarity Triangle\cite{PDG2022}.
One of its angles is \mbox{$\gamma	\equiv 	\arg\left[-(\Vud\Vubs)/(\Vcd\Vcbs)\right]$}.
It can be determined either with direct measurements or indirectly through global fits of the CKM parameters,  assuming unitarity.
Any discrepancy between direct and indirect determinations
would be a hint of physics beyond the SM.
It is possible to directly measure $\gamma$ with processes dominated by tree-level Feynman diagrams, exploiting the interference between the $\bquark \to \cquark$ and $\bquark \to \uquark$ transitions. 
This is usually done by studying $B^{\pm}\to (D\to f_D) h^\pm$ decays where $f_D$ labels a final state shared by the decays of both \Dz and \Dzb mesons and $h$ stands for a kaon or a pion ($h\in \{\pi,K\}$).
Depending on $f_D$, various methods are possible. 
All of them rely on
combination of decay rates and \CP asymmetries observed in the decay of the \B meson and its charge conjugate decay mode.
Other necessary quantities are related to the decay of the \Dz mesons. 
In \lhcb, they are often assumed as external inputs. 
A crucial parameter is the \textit{coherence factor}, \mbox{$R_{f_\D}\in [0,1]$}, which measures the quantum interference in the decay mode.
\lhcb has recently published a $\gamma$ measurement\cite{gamma1}, which exploits $f_D = \Kmp \pipm\pipm\pimp$ and both Run1 and Run2 data.
The global coherence factor is \mbox{$R_{K3\pi}\approx 0.4$}, but in phase space bins it is larger\cite{CoherenceFactor}.
This analysis uses 4 phase-space bins. They are defined according to a previous \lhcb amplitude analysis\cite{gamma1bins}. 
The final result is \mbox{$\gamma = \left(54.8^{+6.0+0.6+6.7}_{-5.8-0.6-4.3}\right)^\circ$}, where the uncertainties are statistical, systematic, and due to external inputs, respectively.
This is the second most precise determination of $\gamma$.
The external inputs come from model-independent determinations by \cleo-c, \besiii, and \lhcb\cite{gamma1extInputs}.
They are the dominant source of uncertainty, but improvements are expected with the upcoming data.
The just mentioned result is already included in the combination\cite{gammaComb} of all direct measurements of $\gamma$ by \lhcb, whose value is \mbox{$\gamma = \left(63.8^{+3.5}_{-3.7}\right)^\circ$}.
It largely dominates the world-average for this quantity and it is compatible with the indirect determinations by both the \textit{CKMfitter} and the \textit{UTfit} groups\cite{gammaCKMandUTfit}.
The very latest determination\cite{gamma2}  of $\gamma$ by the \lhcb collaboration, not yet included in the combination, concerns the cases \mbox{$f_\D = \Kp\Km\pipm\pimp$} and \mbox{$f_\D = \pip\pim\pipm\pimp$}.
A phase-space integrated analysis is performed for both the final states. 
A binned-analysis, which adds substantial sensitivity to $\gamma$, has been completed for the former case.
The result is: $\gamma = (116^{+12}_{-14})^\circ$.
Both Run1 and Run2 data are considered.
The main source of uncertainty is due to the charm decay parameters taken from a model-dependent amplitude analysis by \lhcb\cite{gamma2extInputs}.
Precision and central value may change after the inclusion of upcoming model-independent measurements.



\section{\CP violation in $\Bs\to\phi\phi$ decays}
The $\Bs\to \phi\phi$ channel is a penguin-mediated decay process.
According to the SM, CPV in this mode is suppressed below the experimental sensitivity of \lhcb.
Hence, any CPV enhancement would be a hint of new physics.
The CPV violation in this mode is usually\cite{PDG2022} quantified in terms of \CP-violating phase, $\phi^{\squark\squarkbar\squark}_s$, and direct \CP-vioaltion parameter $|\lambda|$.
To measure them, an angular analysis is needed to disentangle the three polarisation states of the di-vector final state. 
In the following, they will be labelled as $0$, $\parallel$, and $\perp$.
The SM predicts no dependence of $\phi^{\squark\squarkbar\squark}_s$ and $|\lambda|$ on the polarisation. 
This fact can be experimentally verified assuming independent values of the \CP observables:  $\phi_{s,i}$ and $|\lambda|_i$  with $i\in \{0, \parallel,\perp \}$.
The \lhcb collaboration has recently measured these quantities with Run2 data.
The analysis\cite{Bs2phiphi} involve two steps. 
The first one exploits an invariant-mass fit to statistically subtract the background.
After that, the \CP observables are determined through a simultaneous fit to the \Bs decay time and the helicity angles of the final states. 
The information about the flavour of the \Bs meson at production is obtained by dedicated flavour-tagging algorithms\cite{FT}.
Relevant experimental challenges are the calibration of the decay-time resolution and the mistag probability.
When the \CP observables are allowed to be polarisation dependent, the results of this analysis are:
\mbox{$\phi_{s,0} = -0.18 \pm 0.09~{\rm rad}$}, 
\mbox{$\phi_{s,\parallel} - \phi_{s,0} = 0.12 \pm 0.09~{\rm rad}$}, 
\mbox{$\phi_{s,\perp} - \phi_{s,0} = 0.17 \pm 0.09~{\rm rad}$},
\mbox{$|\lambda|_0 = 1.02 \pm 0.17$},
\mbox{$|\lambda_\perp/\lambda|_0  = 0.97 \pm 0.22$},
\mbox{$|\lambda_\parallel/\lambda|_0  = 0.78 \pm 0.21$},
where the uncertainties are statistical only. This is the first determination of these quantities.
No significant differences between various polarization states is observed.
When the same values of the \CP observables are assumed in all the polarisation states, the results of this analysis can be combined with the corresponding ones, obtained by \lhcb with Run1 data. In conclusion:
$\phi^{\squark\squarkbar\squark}_s = -0.074 \pm 0.069$ and $|\lambda| = 1.009 \pm 0.030$.
This is the most precise measurement of time-dependent \CP asymmetry in any penguin-dominated \B meson decay. 
It agrees with the SM expectation of tiny CPV.
\section{Conclusions}
LHCb Run2 data have been providing remarkable insights in both the charm and beauty CPV sectors.
In the last six months, \lhcb reported the first evidence of CPV concerning the \cquark quark in a single decay channel ($3.8\sigma$), a first search for local CPV in charm multi-body decays,
new measurements of the angle $\gamma$ of the UT, 
and the most precise CPV measurements a in penguin-dominated \B decay.
The precision on all these quantities is expected to improve thanks to the data that will be collected during the  ongoing Run3.
In fact, the instantaneous luminosity will be increased by a factor of five and the removal of the hardware trigger will improve the selection efficiency on hadronic final states.

\end{document}